\documentstyle[11pt,epsfig]{article}
\newcommand{\be}{\begin{equation}}
\newcommand{\bee}{\begin{equation}}
\newcommand{\ee}{\end{equation}}
\newcommand{\beqn}{\begin{eqnarray}}
\newcommand{\eeqn}{\end{eqnarray}}

\newcommand{\ra}{\rightarrow}

\newcommand{\kf}{{\bf k}}
\newcommand{\rf}{{\bf r}}
\newcommand{\rhof}{{\bf \rho}}

\newcommand{\qf}{{\bf q}}

\newcounter{savefig}

\begin{document}
\vspace*{-2cm}
\begin{flushright}
DTP/98/56\\
August 1998

\end{flushright}
\vspace{2cm}
%%%%%%%%%%%%%%%%%%%%%%%%%%%%%%%%%%%%%%%%%%%%%%%%%%%%%%%%%%%%%%%%
\begin{center}
{\LARGE \bf 
 Hard diffractive Photon-Proton Scattering \\ at large t  
\\[2mm]}
\vspace{1cm}
D.Yu.~Ivanov$^{ \$}$, 
M.~W\"usthoff$^\star$
\vspace{1cm}

$^{\$}$
Institute of Mathematics, 630090 Novosibirsk, Russia \\

\vspace{2em}
$^{\star}$ University of Durham, Physics Department\\
South Rd., Durham DH1 3LE, UK

\end{center}   

\vspace{1cm}
\noindent{\bf Abstract:}

We propose to test perturbative QCD in the Regge limit  
by means of diffractive photon scattering
$\gamma p \to \gamma X$ at large $t$ and very high energies
$W^2\gg |t|\gg \Lambda^2_{QCD}$. The helicity amplitudes of this 
process were calculated using the Lipatov solution of the BFKL 
equation for $t\neq 0$. We found that the perturbatively 
calculated cross section for this process exceeds the
cross section for $J/\Psi$ photoproduction assuming similar 
kinematics.

\vspace*{\fill}
\eject
\newpage
\section{Introduction}\setcounter{equation}{0}
In the quest for more and better tests of pQCD in the Regge limit,
better known as BFKL \cite{BFKL}, we suggest hard diffractive photon-proton
scattering at large momentum transfer $t$. This process is closely related to 
the diffractive production of vector mesons at large $t$
\cite{ForRys,BFLW,I} where
the vector meson in the final state takes the place of the photon.
Although being suppressed by an extra $\alpha_{em}$ diffractive 
$\gamma p$-scattering has the
great advantage of being completely calculable. No phenomenological
input in terms of a vector meson wave function is needed. The signature
of this process is also very clear. The photoproduced 
photon is scattered into the backward
region of the detector at a very low angle. The transverse momentum 
transferred from the photon is balanced by a jet in the forward region
which does not need to be resolved. Important is the very large rapidity
gap between the photon in the backward and the hadronic system in the 
forward region. The idea to use this process as test for pQCD was
already discussed in \cite{GI}. 

A very precise definition of the cross section is not necessary, since we 
are interested in a rough estimate. Indeed, since BFKL enters on the 
level of the amplitude the resulting enhancement is very large and therefore
the theoretical uncertainty as well. 
We work with the leading order BFKL-solution
which is by now known to receive strong NLO-corrections \cite{FadLip}.
Already a reduction by factor of 1/2 in the LO-BFKL-kernel leads to an 
order of magnitude reduction in the elastic cross section. Another
source of uncertainty is the influence of nonperturbative effects despite
the fact that we require a large momentum transfer $t$.
On the proton side it has been proven \cite{MT,BFLRW} that the large 
$t$-Pomeron-exchange factorizes in the sense that it directly couples to
partons. The difference in the coupling to quarks and gluons
is only a trivial colour factor and the corresponding parton distribution
can be taken from any conventional LO-pdf. As we already pointed out
we are at the present only interested in a first rough estimate of the
pQCD predictions and thus focus on the elastic photon-quark scattering.
For comparison we include in our numerical analysis the diffractive
production of $J/\Psi$ and find that both cross sections are close
in magnitude. The elastic $\gamma q$-cross section, surprisingly, exceeds
the cross section for diffractive $J/\Psi$-production.
We also perform  a Vector Dominance Model
(VDM) inspired estimate for the same cross sections which shows that with 
BFKL the $\gamma q$-cross section for large $t$ can hardly be matched
with the VDM-result at low $t$. At this point, of course, we are most curious
to see what the corresponding measurement will give. If the cross section
is as high as predicted by pQCD, enough events will be recorded (or have been
recorded) to work out an experimental cross section. We will already
learn a lot from the mere presence or absence of events.

In the technical part of the paper we extend the successful 
concept of the photon wave function \cite{Mue,NZ} to include nonzero
momentum transfer. We then convolute the corresponding expression
for the photon wave function directly
in impact parameter space with the conformal
eigenfunction of the nonforward BFKL-solution \cite{Lip}. 
This part is presented
in a rather detailed way to illustrate some of the techniques which might
be useful in other related cases, for it seems most appropriate in dealing
with integrals of two dimensional conformal field theories.
The impact factor describing the coupling of two gluons via a quark-box
to the photons has effectively been already calculated in QED in the context of
photon-photon transition \cite{CW,LF}, only colour factors needed to be added
in the case of QCD.
Since for our calculation we need a somewhat different form 
of the impact factor  than found in the literature, 
we have reconsidered this calculation using slightly different methods.

The paper contains three technical section which deal with the derivation
of the $\gamma q$-cross section. The following section is devoted to the 
generalization of the photon wave function to include momentum transfer, 
in section 3 we explain the convolution of the wave function with the conformal
BFKL-eigenfunctions and in section 4 we collect all pieces to derive
the final expression for the cross section. In section 5 we present
the numerical results and conclude with section 6.  

\section{Photon Wave Function}\setcounter{equation}{0}
In the standard case of
deep inelastic scattering the photon appears only in
the initial state and its wave function \cite{Mue,NZ}
can be formulated in the simplifying photon-proton CMS. 
In this frame the momentum  of the photon has only longitudinal
components and no transverse. When we consider, however, photon
elastic scattering we have two photons, one in the initial state and the 
second in the final state of the process. We can still choose a frame
in which one of the photons has only longitudinal components, but
the second photon receives transverse components due to the momentum
transfer $q$. We therefore have to generalize the photon wave function
description to also include transverse components.
We use the standard notation for deep inelastic scattering such 
as $Q'$ (final state photon) and $p$ (proton). 
The light cone vectors vectors $Q'$
and $p$  define the CMS we work in. For our discussion
it is easier to assume that the outgoing
photon which includes the momentum transfer $q$ lies along the z-axis
of our system and the incoming photon with the momentum $Q$ has 
components in the transverse plane, i.e. $Q=Q'+ q$. 
The corresponding polarization vector for the incoming photon reads
\be\label{pol1} 
\epsilon(\pm)_\mu \;=\;{\epsilon(\pm)_\perp}_\mu\;-\;
\frac{q_\perp \cdot \epsilon(\pm)_\perp}{p\cdot Q}\;p_\mu
\ee
with two helicity states $(\pm)$ in the transverse plane of
our frame:
\be\label{pol2}
{\epsilon_\perp(\pm)}\;=\;\frac{1}{\sqrt{2}} (0,1,\pm i,0)
\ee
\begin{figure}[t]
\begin{center}
\epsfig{file=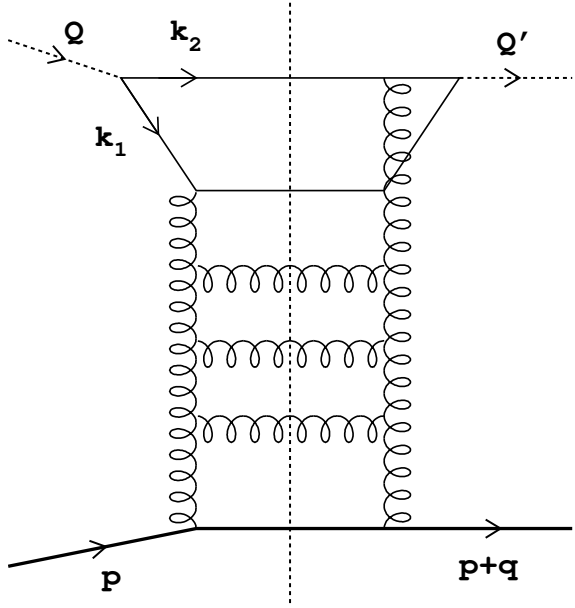,height=10cm}
\caption{Diagrammatical representation of elastic $\gamma q$-scattering
with BFKL-ladder exchange.}
\end{center}
\end{figure}

The photon couples to a quark-antiquark pair with the momenta
$k_1$ and $k_2$, respectively. Using Sudakov decomposition we 
write the momenta as ($s=2Q'\cdot p=4 Q'_0 p_0$) 
\beqn
q&=&\beta_q \;p+q_\perp \nonumber \\
k_1&=& \alpha \;Q' + \left(\beta_q+\frac{k_{2_\perp}^2}
{(1-\alpha) \;s}
\right)\;p + k_{1_\perp}\\
k_2&=&(1-\alpha)\; Q' - \frac{k_{2_\perp}^2}{(1-\alpha)\; s}\; p 
+ k_{2_\perp}\;\;.
\nonumber
\eeqn
One of these quark momenta is offshell depending on where the t-channel
gluon couples. In the case discussed here it is 
$k_1$ as indicated in Fig.1. $k_1$ can be made onshell by
adding a component with respect to $p$. We make use of this trick
in order to break up the trace of the quark loop by inserting quark spinors
with onshell momenta. This can be done without 
effecting the final result because the adjacent gluon coupling has
as leading component a $\rlap / p$ which cancels our proposed 
modification in the trace. In the trace we therefore 
substitute $k_1$ by $\tilde{k}_1$:
\be
\tilde{k}_1\;=\;\alpha \;Q' - \frac{k_{1_\perp}^2}{\alpha\; s} \;p 
+ k_{1_\perp}\\
\ee
The denominator of our wave function contains of course the original virtuality
$k_1^2$. In the following we will change the transverse components
for the momenta from Minkowski space to Euclidean, i.e $k_\perp \rightarrow
\kf $ with $k_\perp^2=-\kf^2$, and despite the fact that 
the photon virtuality is negative
we will use the convention that $Q^2=|Q^2|=-\beta_q s+\qf^2$.

After these preliminaries it is rather straightforward to introduce 
the photon wave function ($h=\pm\;$ for the quark helicity, 
$h_\gamma=\pm$ for the photon helicity):
\beqn\label{wave1}
\Psi_{(h_\gamma,h)}(\kf_1,\kf_2,\alpha)&=&e\;\frac{\bar{u}(k_2,h)\; 
\rlap / \epsilon(h_\gamma)\;
u(\tilde{k}_1,h) \sqrt{\alpha(1-\alpha)}}
 {[\kf_1-\alpha\qf]^2+\alpha(1-\alpha)Q^2}
\;(2 \pi)^2 \delta^2(\qf-\kf_1-\kf_2)
\eeqn
In the expression above we have made use of the relation $v(\tilde{k}_1,-h)
=u(\tilde{k}_1,h)$ and we have already incorporated a factor $\alpha $ from 
the gluon vertex and a phase space contribution $1/\sqrt{\alpha(1-\alpha)}$. 
The propagator 
itself has the form $|k_1^2|=([\kf_1-\alpha\qf]^2+\alpha(1-\alpha)Q^2)/
(1-\alpha)$. In the chiral representation the spinors are of the following 
form (in complex notation for $\kf_1$):
\beqn\label{spin}
u(\tilde{k}_1,+)&=&\left(\frac{\kf_1^*}{\sqrt{2 \alpha Q'_0}}\;,\;
\sqrt{2 \alpha Q'_0}\;,\;0\;,\;0\right) \\
u(\tilde{k}_1,-)&=&\left(0\;,\;0,\;\sqrt{2 \alpha Q'_0}\;,\;\frac{-\kf_1}
{\sqrt{2 \alpha Q'_0}}\right) \nonumber
\eeqn
and similar expressions for $u(k_2,h)$ by substituting $\kf_1$ by $\kf_2$
and $\alpha$ by $1-\alpha$. 

Eq.(\ref{wave1}) together with (\ref{pol1}) and (\ref{spin}) 
yields the following expressions for the wave function 
(for better readability we present
all helicity states explicitly):
\beqn \label{wave2}
\Psi_{(+,+)}(\kf_1,\kf_2,\alpha)&=&-
\frac{\sqrt{2}\;e\;\alpha\; [\kf_2-(1-\alpha)\qf]}
 {[\kf_2-(1-\alpha)\qf]^2+\alpha(1-\alpha)Q^2} 
\;(2 \pi)^2 \delta^2(\qf-\kf_1-\kf_2)\nonumber \\
\Psi_{(+,-)}(\kf_1,\kf_2,\alpha)&=&-
\frac{\sqrt{2}\;e\;(1-\alpha)\; [\kf_1-\alpha\qf]}
 {[\kf_1-\alpha\qf]^2+\alpha(1-\alpha)Q^2}
\;(2 \pi)^2 \delta^2(\qf-\kf_1-\kf_2)    \\
\Psi_{(-,+)}(\kf_1,\kf_2,\alpha)&=&-
\frac{\sqrt{2}\;e\;(1-\alpha)\; [\kf_1^*-\alpha\qf^*]}
 {[\kf_1-\alpha\qf]^2+\alpha(1-\alpha)Q^2}
\;(2 \pi)^2 \delta^2(\qf-\kf_1-\kf_2)\nonumber \\
\Psi_{(-,-)}(\kf_1,\kf_2,\alpha)&=&-
\frac{\sqrt{2}\;e\;\alpha\; [\kf_2^*-(1-\alpha)\qf^*]}
 {[\kf_2-(1-\alpha)\qf]^2+\alpha(1-\alpha)Q^2}
\;(2 \pi)^2 \delta^2(\qf-\kf_1-\kf_2) \nonumber
\eeqn
One could of course apply momentum conservation
$\kf_2=\qf-\kf_1$ and rewrite
all equations in terms of $\kf_1$ only. But it is the form presented in
(\ref{wave2}) which emerges
first and still exhibits the fact that  
the same factor $\alpha(1-\alpha)$ appears
in front of $\qf$ in the numerator of all expressions. The latter
is a result of the longitudinal contribution $p_\mu$ in the polarization vector
(\ref{pol1}) which is the same for all quark helicities.

We transform into impact parameter space by taking the Fourier transform
with respect to $\kf_1$ and $\kf_2$
\be 
\int \frac{d^2 \kf_1 d^2 \kf_2}{(2 \pi)^4}\; \Psi(\kf_1,\kf_2,\alpha)\;
\mbox{e}^{i \kf_1 \cdot \rf_1}\; \mbox{e}^{i \kf_2 \cdot \rf_2}
\;=\; \Psi(\rf_1,\rf_2,\alpha)
\ee
and arrive at 
\beqn\label{wavevirtual}
\Psi_{(+,+)}(\rf_1,\rf_2,\alpha)&=& \frac{\sqrt{2}\; i\;e}{2 \pi}\;\alpha \;
\sqrt{\alpha(1-\alpha) Q^2}\;K_1(\sqrt{\alpha(1-\alpha) Q^2}\;|\rf|)\;
\frac{\rf}{|\rf|}\;
\mbox{e}^{i \alpha \qf \cdot \rf_1}\;\mbox{e}^{i (1-\alpha) \qf \cdot \rf_2}
\nonumber \\
\Psi_{(+,-)}(\rf_1,\rf_2,\alpha)&=& \frac{\sqrt{2}\;i\;e}{2 \pi}\;(1-\alpha)\;
\sqrt{\alpha(1-\alpha) Q^2}\;K_1(\sqrt{\alpha(1-\alpha) Q^2}\;|\rf|)\;
\frac{\rf}{|\rf|}\;
\mbox{e}^{i \alpha \qf \cdot \rf_1}\;\mbox{e}^{i (1-\alpha) \qf \cdot \rf_2}
\nonumber \\
\Psi_{(-,+)}(\rf_1,\rf_2,\alpha)&=& \frac{\sqrt{2}\;i\;e}{2 \pi}\;(1-\alpha)\;
\sqrt{\alpha(1-\alpha) Q^2}\;K_1(\sqrt{\alpha(1-\alpha) Q^2}\;|\rf|)\;
\frac{\rf^*}{|\rf|}\;
\mbox{e}^{i \alpha \qf \cdot \rf_1}\;\mbox{e}^{i (1-\alpha) \qf \cdot \rf_2}
\nonumber \\
\Psi_{(-,-)}(\rf_1,\rf_2,\alpha)&=& \frac{\sqrt{2}\;i\;e}{2 \pi}\;\alpha\;
\sqrt{\alpha(1-\alpha) Q^2}\;K_1(\sqrt{\alpha(1-\alpha) Q^2}\;|\rf|)\;
\frac{\rf^*}{|\rf|}\;
\mbox{e}^{i \alpha \qf \cdot \rf_1}\;\mbox{e}^{i (1-\alpha) \qf \cdot \rf_2}
\eeqn
with $\rf=\rf_1-\rf_2$ and $K_1$ being the MacDonald (Bessel) function. 
The whole $\qf$-dependence is shifted into phase factors, and
the rest is the same as for the 'forward' case, i.e. $\qf=0$.

One could now go ahead and work with the above wave function for virtual
photons, i.e study the scattering of virtual photons into virtual photons. 
At this point we would like to keep the analysis a bit simpler and
consider only real photons. To this end we take $Q^2 \rightarrow 0$
and eq.(\ref{wavevirtual}) reduces to
\beqn
\Psi_{(+,+)}(\rf_1,\rf_2,\alpha)&=& \frac{\sqrt{2}\;i\;e}{2 \pi}\;\alpha \;
\frac{\rf}{|\rf|^2}\;
\mbox{e}^{i \alpha \qf \cdot \rf_1}\;\mbox{e}^{i (1-\alpha) \qf \cdot \rf_2}
\nonumber \\
\Psi_{(+,-)}(\rf_1,\rf_2,\alpha)&=& \frac{\sqrt{2}\;i\;e}{2 \pi}\;(1-\alpha)\;
\frac{\rf}{|\rf|^2}\;
\mbox{e}^{i \alpha \qf \cdot \rf_1}\;\mbox{e}^{i (1-\alpha) \qf \cdot \rf_2}
\nonumber \\
\Psi_{(-,+)}(\rf_1,\rf_2,\alpha)&=& \frac{\sqrt{2}\;i\;e}{2 \pi}\;(1-\alpha)\;
\frac{\rf^*}{|\rf|^2}\;
\mbox{e}^{i \alpha \qf \cdot \rf_1}\;\mbox{e}^{i (1-\alpha) \qf \cdot \rf_2}
\\
\Psi_{(-,-)}(\rf_1,\rf_2,\alpha)&=& \frac{\sqrt{2}\;i\;e}{2 \pi}\;\alpha \;
\frac{\rf^*}{|\rf|^2}\;
\mbox{e}^{i \alpha \qf \cdot \rf_1}\;\mbox{e}^{i (1-\alpha) \qf \cdot \rf_2}
\;\;.\nonumber 
\eeqn

In the next step we have to consider the wave function for the outgoing photon
$\Psi^*$. It is basically the complex conjugate of the previous expression
except for the phase factors which are absent because the momentum for
the outgoing photon is simply $Q'$, i.e. $q=0$: 
\beqn
\Psi^*_{(+,+)}(\rf_1,\rf_2,\alpha)&=&-\; \frac{\sqrt{2}\;i\;e}{2 \pi}\;
\alpha \;\frac{\rf^*}{|\rf|^2}\;
\nonumber \\
\Psi^*_{(+,-)}(\rf_1,\rf_2,\alpha)&=&-\;\frac{\sqrt{2}\;i\;e}{2 \pi}\;
(1-\alpha) \;\frac{\rf^*}{|\rf|^2}\;
\nonumber \\
\Psi^*_{(-,+)}(\rf_1,\rf_2,\alpha)&=&-\; \frac{\sqrt{2}\;i\;e}{2 \pi}\;
(1-\alpha) \;\frac{\rf}{|\rf|^2}\;
\\
\Psi^*_{(-,-)}(\rf_1,\rf_2,\alpha)&=&-\;\frac{\sqrt{2}\;i\;e}{2 \pi}\;
\alpha \;\frac{\rf}{|\rf|^2}\;
\;\;.\nonumber 
\eeqn

The impact factor is essential the product of the two sets of wave functions,
since the imaginary part of our amplitude dominates and all intermediate
quarks are onshell. With regard to the two t-channel gluons
located at $\rhof_1$ and $\rhof_2$
we have to make sure that each of the gluons couples  
to each of the quarks (g is the strong coupling constant):
\beqn
\Phi_{(h_\gamma,h^*_\gamma)}(\rhof_1,\rhof_2)&=&g^2\;\sum_{h=\;\pm}
\int_0^1 d \alpha\;\int d^2 \rf_1 d^2 \rf_2\;
\Psi_{(h_\gamma,h)}(\rf_1,\rf_2,\alpha)\;
\Psi^*_{(h^*_\gamma,h)}(\rf_1,\rf_2,\alpha)\\
&&\hspace{2cm}[\delta^2(\rf_1-\rhof_1)-\delta^2(\rf_2-\rhof_1)]\;
[\delta^2(\rf_1-\rhof_2)-\delta^2(\rf_2-\rhof_2)]\nonumber\;\;.
\eeqn
The normalization will be adjusted later in the full expression for the 
amplitude. Spelling out the previous expression we find:
\beqn\label{impact}
\Phi_{(+,+)}(\rhof_1,\rhof_2)&=& \nonumber\\
\Phi_{(-,-)}(\rhof_1,\rhof_2)&=&
\frac{e^2g^2}{2 \pi^2}\;
\int_0^1 d \alpha\;\int d^2 \rf_1 d^2 \rf_2\;[\alpha^2+(1-\alpha)^2]\;
\frac{1}{|\rf|^2}\;
\mbox{e}^{i \alpha \qf \cdot \rf_1}\;\mbox{e}^{i (1-\alpha) \qf \cdot \rf_2}\;
\nonumber\\ 
&&\hspace{1cm}[\delta^2(\rf_1-\rhof_1)-\delta^2(\rf_2-\rhof_1)]\;
[\delta^2(\rf_1-\rhof_2)-\delta^2(\rf_2-\rhof_2)]\\
\Phi_{(+,-)}(\rhof_1,\rhof_2)&=&\frac{e^2g^2}{\pi^2}\;
\int_0^1 d \alpha\;\int d^2 \rf_1 d^2 \rf_2\;\alpha(1-\alpha)\;
\frac{\rf^2}{|\rf|^4}\;
\mbox{e}^{i \alpha \qf \cdot \rf_1}\;\mbox{e}^{i (1-\alpha) \qf \cdot \rf_2}\;
\nonumber\\
&&\hspace{1cm}[\delta^2(\rf_1-\rhof_1)-\delta^2(\rf_2-\rhof_1)]\;
[\delta^2(\rf_1-\rhof_2)-\delta^2(\rf_2-\rhof_2)]\nonumber \\
\Phi_{(-,+)}(\rhof_1,\rhof_2)&=&\frac{e^2g^2}{\pi^2}\;
\int_0^1 d \alpha\;\int d^2 \rf_1 d^2 \rf_2\;\alpha(1-\alpha)\;
\frac{{\rf^*}^2}{|\rf|^4}\;
\mbox{e}^{i \alpha \qf \cdot \rf_1}\;\mbox{e}^{i (1-\alpha) \qf \cdot \rf_2}\;
\nonumber\\
&&\hspace{1cm}[\delta^2(\rf_1-\rhof_1)-\delta^2(\rf_2-\rhof_1)]\;
[\delta^2(\rf_1-\rhof_2)-\delta^2(\rf_2-\rhof_2)]\nonumber
\eeqn
$\Phi_{(+,+)}$ and $\Phi_{(-,-)}$ are contributions without helicity flip and
$\Phi_{(+,-)}$ and $\Phi_{(-,+)}$ are those with flip. It is not apparent
from eq.(\ref{impact}) that the two helicity flip contributions are the same, 
but the following calculation will show that they coincide as one may
expect on general grounds. In the end it is enough to deal with
a single helicity flip and a single helicity non-flip amplitude.

\section{Projection on Conformal Eigenstates}\setcounter{equation}{0}
In this section we exploit the properties of conformal invariance of
the BFKL-equation \cite{Lip} which can be solved in terms of the conformal
covariant eigenfunctions:
\be\label{eigen}
E^\nu(\rhof_{10},\rhof_{20})\;=\;\left|\frac{\rhof_{12}}{\rhof_{10}\rhof_{20}}
\right|^{1+2 i \nu}
\ee
where $\nu$ is the conformal weight. We have ignored the conformal spin $n$
which is required to form a complete set. In practice, though, any
contribution with $n\ne 0$ gives a subleading contribution at high energies.
If wished to do so, the following calculation can be generalized to include
$n$. 

In projecting the impact factor on the eigenfunction we have to perform
the following integration ($\rhof_{10}=\rhof_1-\rhof_0$, etc.):
\be
\int d^2 \rhof_1  d^2 \rhof_2\; \Phi_{(h_\gamma,h^*_\gamma)}(\rhof_1,\rhof_2)\;
E^\nu(\rhof_{10},\rhof_{20})\;\;.
\ee
We realize that in two terms of eq.(\ref{impact}) 
the $\delta$-function forces $\rhof_1$ and $\rhof_2$ into one point
which results
into a vanishing contribution due to a zero in $E^\nu$
( $Re[1+2i\nu]$ has to be kept positive). The second observation
is the symmetry in $\rhof_1$ and $\rho_2$ which allows to write the whole
amplitude in one term multiplied by 2 (non-flip here): 
\beqn 
\frac{e^2g^2}{\pi^2}\;
\int_0^1 d \alpha\;[\alpha^2+(1-\alpha)^2]\;\int d^2 \rf_1 d^2 \rf_2\;
\frac{1}{|\rf|^2}\;
\mbox{e}^{i \alpha \qf \cdot \rf_1}\;\mbox{e}^{i (1-\alpha) \qf \cdot \rf_2}\;
\left|\frac{\rf}{(\rf_1-\rhof_0)(\rf_2-\rhof_0)}\right|^{1+2 i \nu}\;\;.
\eeqn
We will shift $\rf_1$ and $\rf_2$ by $\rho_0$ generating an overall
phase factor and then substitute $\rf_1$ by $\rf=\rf_1-\rf_2$:
\beqn\label{rho_0} 
\frac{e^2g^2}{\pi^2}\;\mbox{e}^{i \qf \cdot \rhof_0}
\int_0^1 d \alpha\;[\alpha^2+(1-\alpha)^2]\;\int d^2 \rf\; d^2 \rf_2\;
\mbox{e}^{i \alpha \qf \cdot \rf}\;\mbox{e}^{i \qf \cdot \rf_2}\;
|\rf|^{-1+2i\nu}\;|(\rf+\rf_2)\;\rf_2|^{-1-2 i \nu}\;\;.
\eeqn
The overall phase factor will be ignored for a while and reconsidered
at the end of this section.
We carry on in our calculation and shift $\rf_2$ by $-\alpha\; \rf$ 
eliminating the phase factor that depends on $\rf$ 
\beqn 
\frac{e^2g^2}{\pi^2}\;
\int_0^1 d \alpha\;[\alpha^2+(1-\alpha)^2]\;\int d^2 \rf\; d^2 \rf_2\;
\;\mbox{e}^{i \qf \cdot \rf_2}\;
|\rf|^{-1+2i\nu}\;|[\rf_2+(1-\alpha)\rf]\;[\rf_2-\alpha\rf]|^{-1-2 i \nu}\;\;.
\eeqn
We then switch from the conventional representation of the transverse 
vectors to the corresponding complex notation, i.e from 
$\rf=(r_1,r_2)$ to $a=r_1+i\;r_2$ and $b=r_1-i \;r_2$,
and make use of the freedom to rotate our system such that $\qf$ is real:
\beqn 
&-&\frac{e^2g^2}{4\pi^2}\;
\int_0^1 d \alpha\;[\alpha^2+(1-\alpha)^2]\;\int d a \;d b\; d a_2\; d b_2 \;
\;\mbox{e}^{i q/2 (a_2+b_2)}\;(a\;b)^{-1/2+i\nu}\\
&&\hspace{2cm}\{ [a_2+(1-\alpha)a]\;[b_2+(1-\alpha)b]\;
[a_2-\alpha\;a][b_2-\alpha\;b]\}^{-1/2- i \nu}\;\;.\nonumber
\eeqn
In what follows we rescale $a$ and $b$ by $a_2$ and $b_2$, respectively, and
thus factorize the integration in $a,b$ and $a_2,b_2$:
\beqn 
&-&\frac{e^2g^2}{4\pi^2}\;
\int_0^1 d \alpha\;[\alpha^2+(1-\alpha)^2]\;\int d a_2 \;d b_2\; 
\;\mbox{e}^{i q/2 (a_2+b_2)}\;(a_2\;b_2)^{-1/2-i\nu} \\
&&\hspace{0cm}\int d a \;d b\;\;(a\;b)^{-1/2+i\nu}\;
\{ [1+(1-\alpha)a]\;[1+(1-\alpha)b]\;
[1-\alpha\;a][1-\alpha\;b]\}^{-1/2- i \nu}\;\;.\nonumber
\eeqn
We now rescale $a_2$ and $b_2$ by $2 i/q$
\beqn \label{rescaled}
&&\frac{e^2g^2}{4\pi^2}\;\left(\frac{2}{q}\right)^{1-2i\nu}\;
\int_0^1 d \alpha\;[\alpha^2+(1-\alpha)^2]\;\int d a_2 \;d b_2\; 
\;\mbox{e}^{-(a_2+b_2)}\;(-a_2\;b_2)^{-1/2-i\nu} \\
&&\hspace{0cm}\int d a \;d b\;\;(a\;b)^{-1/2+i\nu}\;
\{ [1+(1-\alpha)a]\;[1+(1-\alpha)b]\;
[1-\alpha\;a][1-\alpha\;b]\}^{-1/2- i \nu}\nonumber
\eeqn
and note the minus sign in the factor $(-a_2\;b_2)^{-1/2-i\nu}$.

At this point we have to consider the analytic structure of our integrand
and the proper path of integration. We focus on the $a_2$-integration
by keeping $b_2$ fixed, quasi as a parameter. With the minus sign there
is a cut in the complex $a_2$ plane to right when $b_2>0$ or to the left
when $b_2<0$. Since the exponent also has a minus sign we would like to
shift the integration contour to the right. A nonzero contribution only
arises, if a singularity is encountered, i.e. if $b_2$ is
positive. Closing the contour around the cut
means we have to take the discontinuity along the cut and
then integrate in $a_2$ over the positive axis:
\beqn\label{ab1}
&&\int d a_2 \;d b_2\; 
\mbox{e}^{-(a_2+b_2)}\;(-a_2\;b_2)^{-1/2-i\nu}\nonumber\\
&=&
-\;2 i\; \sin[(-1/2-i\nu)\pi] 
\;\int_0^{\infty} d a_2 \; a_2^{-1/2-i\nu}\;\mbox{e}^{-a_2}
\;\int_0^{\infty} d b_2 \; b_2^{-1/2-i\nu}\;\mbox{e}^{-b_2}\\
&=&-\;2 i\; \sin[(-1/2-i\nu)\pi] \;\Gamma^2(1/2-i \nu)\;\;.\nonumber
\eeqn 
The $-\;2 i\; \sin[(-1/2-i\nu)\pi]$ is the result of taking the discontinuity.

In a similar way the integration over $a$ and $b$ can be performed. 
We again keep $b$ fixed and consider the integration over $a$. The integrand
is convergent even without the presence of an exponent. 
There will be a nonzero
contribution only if the integration contour runs between two cuts,
those two cuts which emerge on both sides of the complex $a$-plane when
$-1/(1-\alpha)<b<1/\alpha$. We have to face, however, another cut due to the
factor $(a\;b)^{-1/2+i\nu}$ in eq.(\ref{rescaled}). This lies to the right
or to the left depending on the sign of $b$. Therefore we have to consider
two cases $-1/(1-\alpha)<b<0$ and $0<b<1/\alpha$ and close the contour
to that side which has only one cut:
\beqn
&&\int d a \;d b\;\;(a\;b)^{-1/2+i\nu}\;
\{ [1+(1-\alpha)a]\;[1+(1-\alpha)b]\;
[1-\alpha\;a][1-\alpha\;b]\}^{-1/2- i \nu}\nonumber\\ 
&=&2 i\; \sin[(-1/2-i\nu)\pi]\;
\left\{ \int_{-1/(1-\alpha)}^0 d b \;\;\int_{-\infty}^{-1/(1-\alpha)} d a
+\int_0^{1/\alpha} d b \;\;\int^{\infty}_{1/\alpha} d a \right\}\\
&&(a\;b)^{-1/2+i\nu}\;\{- [1+(1-\alpha)a]\;[1+(1-\alpha)b]\;
[1-\alpha\;a][1-\alpha\;b]\}^{-1/2- i \nu}\nonumber \\
&=&4i\;\; \sin[(-1/2-i\nu)\pi]\;
[\alpha(1-\alpha)]^{-1/2-i\nu}\;\int_0^1 d b\;b^{-1/2+i\nu}\;
\left\{[1-b][1+\frac{\alpha}{1-\alpha}b]\right\}^{-1/2-i\nu}\;\nonumber \\
&&\hspace{5cm}\int_0^1 d a\;a^{-1/2+i\nu}\;
\left\{[1-a][1+\frac{1-\alpha}{\alpha}a]\right\}^{-1/2-i\nu}\nonumber\;\;.
\eeqn
where we have made use of the symmetry between $\alpha$ and $1-\alpha$.
The integrals above can readily be expressed in terms of Hypergeometric
functions:
\beqn\label{ab}
&=&4i\;\;\sin[(-1/2-i\nu)\pi]\;[\alpha(1-\alpha)]^{-1/2-i\nu}\;
\;\Gamma^2(1/2-i\nu)\;\Gamma^2(1/2+i\nu)\nonumber \\
&&\hspace{1cm}\mbox{$_2$}F_1\left(1/2+i\nu,1/2+i\nu;1;
\frac{\alpha}{\alpha-1}\right)\;
\mbox{$_2$}F_1\left(1/2+i\nu,1/2+i\nu;1;\frac{\alpha-1}{\alpha}\right)
\nonumber \\
&=&4\i\; \sin[(-1/2-i\nu)\pi]\;\;\frac{\pi^2}{\sin^2[(1/2-i\nu)\pi]}\\
&&\hspace{1cm}\mbox{$_2$}F_1\left(1/2-i\nu,1/2+i\nu;1;\alpha\right)\;
\mbox{$_2$}F_1\left(1/2-i\nu,1/2+i\nu;1;1-\alpha\right)
\nonumber \;\;.
\eeqn
Putting the results in eq.(\ref{ab1}) and (\ref{ab}) back into 
eq.(\ref{rescaled}) we get:
\beqn\label{nonflip} 
&&2\;e^2g^2\;\left(\frac{2}{q}\right)^{1-2i\nu}\;\;\Gamma^2(1/2-i \nu)\;
\int_0^1 d \alpha\;[1-2(1-\alpha)+2(1-\alpha)^2]\\
&&\hspace{2.5cm}\mbox{$_2$}F_1\left(1/2-i\nu,1/2+i\nu;1;1-\alpha\right)\;
\mbox{$_2$}F_1\left(1/2-i\nu,1/2+i\nu;1;\alpha\right)
\nonumber \;\;.
\eeqn

For the flip amplitude we have to proceed in a very similar way and find:
\beqn \label{flip}
&&-\;2\;e^2g^2\;\left(\frac{2}{q}\right)^{1-2i\nu}\;\;\Gamma^2(1/2-i \nu)\;
(1/4+\nu^2)\;\int_0^1 d \alpha\;\alpha(1-\alpha)\\
&&\hspace{2.5cm}\mbox{$_2$}F_1\left(1/2-i\nu,1/2+i\nu;2;1-\alpha\right)\;
\mbox{$_2$}F_1\left(1/2-i\nu,1/2+i\nu;2;\alpha\right)
\nonumber 
\eeqn
which is the same for the $(+,-)$- or the $(-,+)$-amplitude.

The further calculation is rather tedious but straight forward. The 
main point is to expand the first Hypergeometric function, perform the 
$\alpha$-integration and then reduce the remaining series. The final result
takes on the following form:
\be\label{nonflip1}
e^2g^2\;\left(\frac{2}{q}\right)^{1-2i\nu}\;\frac{\Gamma(1/2-i\nu)}
{\Gamma(1/2+i\nu)}\;\frac{\pi^2}{4}\;\frac{11/4+3\nu^2}{1+\nu^2}
\;\frac{\tanh(\pi \nu)}{\pi \nu}
\ee
for the non-flip amplitude and 
\be\label{flip1}
e^2g^2\;\left(\frac{2}{q}\right)^{1-2i\nu}\;\frac{\Gamma(1/2-i\nu)}
{\Gamma(1/2+i\nu)}\;\frac{\pi^2}{4}\;\frac{1/4+3\nu^2}{1+\nu^2}
\;\frac{\tanh(\pi \nu)}{\pi \nu}
\ee
for the flip amplitude.

We are left with the calculation of the other end of the BFKL-Pomeron, i.e.
the coupling to a quark(gluon) in the proton. 
We follow the consideration in ref.\cite{MT}
where it was noted that a simple projection of the eigenfunction (\ref{eigen})
to a single quark line would give zero. In reality the quark
is accompanied by a bunch of particles
with opposite colour far away in impact parameter space. 
Let us place the quark 
in impact parameter space  at $\rf'_1$ and the opposite colour charge
at $\rf'_2$. In the limit $|\rf'_2| \gg |\rf'_1|$
the eigenfunction (\ref{eigen}) reduces to
\be
 E^\nu(\rhof'_{10},\rhof'_2\ra \infty)\sim \left|\frac{1}{\rhof'_{10}}
\right|^{1-2i\nu}\;\;.
\ee 
The momentum coming from the photon is transferred to the quark
but momentum conservation is imposed later on by doing the final 
$\rhof_0$-integration.
At the moment we give the quark a general momentum kick $\qf'$ which leads
to a phase factor e$^{i \qf' \cdot \rf'_1}$. The integration over
$\rf'_2$ is factored off and only the $\rf'_1$-integration remains:
\beqn\label{line}
&&2 g^2\;\int \frac{d^2 \rf'_1}{(2\pi)^2}\; \left|\frac{1}{\rf'_1-\rhof_0}
\right|^{1-2i\nu}\;\mbox{e}^{-i \qf' \cdot \rf'_1} \\
&=&g^2\;\left(\frac{2}{q}\right)^{1+2i\nu}\;\frac{\Gamma(1/2+i\nu)}
{\Gamma(1/2-i\nu)}\;\frac{1}{2\pi}\;\mbox{e}^{-i \qf' \cdot \rhof_0}\nonumber
\eeqn
A factor 2 was added to account for the coupling of both gluons.

We have to recall that we dropped the phase factor e$^{i \qf \cdot \rhof_0}$
in eq.(\ref{rho_0}). In the full amplitude the two phase factors have
to be drawn together and the integration over $\rhof_0$ forces $\qf$ and 
$\qf'$ to be equal:
\be \label{pi2}
\int d^2 \rhof_0 \mbox{e}^{i (\qf-\qf') \cdot \rhof_0}\;=\;
(2\pi)^2\;\delta^2(\qf-\qf')\;\;.
\ee
For us important is the extra factor $(2\pi)^2$ which is generated in
the above equation.

\section{The complete Cross Section}\setcounter{equation}{0}
In order to write down the cross section we have to find the complete
expression for the amplitude. Cross section and amplitude are linked
through the relation ($t=\qf^2$)
\beqn
\frac{d \sigma}{d t}(\gamma q \ra \gamma q)&=&\frac{1}{2}\;\frac{
|A_{(+,+)}|^2+|A_{(-,-)}|^2+|A_{(-,+)}|^2
+|A_{(+,-)}|^2}{16 \pi s^2}\;\\
&=&\frac{|A_{(+,+)}|^2+|A_{(+,-)}|^2}{16 \pi s^2}\;\nonumber\;\;.
\eeqn

All missing factors not yet being included
are extracted from the 'Born'-diagram: 
6/9 for the light flavour charges, 4/6 for the colour (in the case
of $\gamma q$-scattering), 4 from the
coupling to the lower line, $s/4$ from the Sudakov decomposition,
$1/2 (2\pi)^3$ for the onshell quarks (1/2 because we need the imaginary part
and not the discontinuity) and $1/(2\pi)^8$ from the phase space integral.
All factors related to Fourier transformations have been taken care of, only
the factor $(2\pi)^2$ from (\ref{pi2}) needs to be included.
Compiling all these factors and add them together with expression 
(\ref{nonflip1}),(\ref{flip1}) and (\ref{line}) we finally obtain
for the amplitudes 
\beqn\label{finalnonflip}
A_{(+,+)}&=&i\;\frac{6}{9} \;\alpha_{em}\;\alpha_s^2\;\frac{4 \pi}{3}\;
\frac{s}{|t|}\;\int d\nu\; \frac{\nu^2}{(1/4+\nu^2)^2}\;
\frac{11/4+3\;\nu^2}{1+\nu^2}\;\frac{\tanh(\pi \nu)}{\pi \nu}\;
\left(\frac{s}{|t|}\right)^{\omega(\nu)}
\eeqn
and
\beqn\label{finalflip}
A_{(+,-)}&=&i\;\frac{6}{9} \;\alpha_{em}\;\alpha_s^2\;\frac{4 \pi}{3}\;
\frac{s}{|t|}\;\int d\nu\; \frac{\nu^2}{(1/4+\nu^2)^2}\;
\frac{1/4+\nu^2}{1+\nu^2}\;\frac{\tanh(\pi \nu)}{\pi \nu}\;
\left(\frac{s}{|t|}\right)^{\omega(\nu)}
\eeqn
where 
\be
\omega(\nu)\;=\;\frac{3 \alpha_s}{ \pi}\;[2\;\Psi(1)-\Psi(1/2+i\nu)
-\Psi(1/2-i\nu)] \;\;.
\ee
One can derive the contribution for $\gamma g$-scattering simply by multiplying
the previous expressions by a factor 9/4 (due to colour).

There is a striking similarity between the results here and those found
for the forward jet cross section with azimuthal dependence \cite{BDW}.
The factor $\frac{11/4+3\;\nu^2}{1+\nu^2}$ for the non-flip contribution
is identical to the corresponding integrated contribution to the forward
jet cross section whereas $\frac{1/4+\nu^2}{1+\nu^2}$, the flip result, 
is found for the azimuth dependent part.  

The saddle point approximation for the amplitudes (\ref{finalnonflip})
and (\ref{finalflip}) yields
\be \label{saddlenonflip}
A_{(+,+)}\;=\;11\;i\;\frac{6}{9} \;\alpha_{em}\;\alpha_s^2\;\frac{s}{|t|}
\;\frac{8}{3}\;\left(\frac{\pi}
{7 \zeta(3)\;\eta}\right)^{3/2}\;\mbox{e}^{\eta \ln(4)}
\ee
and
\be\label{saddleflip}
A_{(+,-)}\;=\;i\;\frac{6}{9} \;\alpha_{em}\;\alpha_s^2\;\frac{s}{|t|}
\;\frac{8}{3}\;\left(\frac{\pi}
{7 \zeta(3)\;\eta}\right)^{3/2}\;\mbox{e}^{\eta \ln(4)}
\ee
where $\eta$ is defined as 
\be
\eta\;=\;\frac{6 \alpha_s}{\pi}\;\ln\left(\frac{s}{|t|}\right)
\ee
In this approximation one nicely sees the dominance of the non-flip 
versus the flip amplitude given by the factor of 11 
in eq.(\ref{saddlenonflip}). On the scale we perform our numerical analysis
the difference between the exact result and the saddle point 
approximation is marginally, i.e. less than a factor of 2.

\section{Numerical Results}\setcounter{equation}{0}
Since we are mainly interested in a rough estimate for the cross section,
we will concentrate on elastic photon-quark scattering.
In order to get an impression for the size of the cross section we compare
with a different process, namely diffractive $J/\Psi$-photoproduction.
In addition we use the Vector Dominance Model (VDM) to gain an estimate
of the quasi elastic $\gamma q$ scattering at low $t$, and we perform 
a comparison between the full BFKL solution and the leading order
two-gluon exchange.

The formulae for the $J/\Psi$-cross section are taken from refs.\cite{ForRys}
and \cite{BFLW}. The leading order two gluon exchange can be more easily 
calculated from the momentum representation of the photon wave function 
(\ref{wave2}) directly. The amplitudes in this case are given by
\beqn
A_{(+,+)}&=&i\;\frac{s}{|t|}\;\alpha_{em}\alpha_s^2
\;\frac{2^6}{3^3}\;\left(\frac{\pi^2}{3}+1\right)\\
A_{(+,-)}&=&i\;\frac{s}{|t|}\;\alpha_{em}\alpha_s^2\;
\frac{2^5}{3^2}\;\;.\nonumber
\eeqn
A VDM-estimate can be achieved by employing the optical theorem
which relates the elastic with the total $\gamma p$-cross section. 
The total $\gamma p$-cross section at the energy of $W=200~GeV$
is approximately $150~{\mu}b$. This leads together with the $t$-slope 
of $10~GeV^{-2}$ to an integrated 
elastic $\gamma p$-cross section of $115~nb$.
From the diffractive production of $\rho^0$'s one knows that for
roughly 11\% of all events the proton dissociates. The resulting value
for the cross section 
divided by 9 (additive quark model) gives an estimate for the 
$\gamma q$-scattering of $1.4~nb$ at $W=200GeV$.
The diffractive slope in $t$ is according to the $\rho^0$-data  
$5.3~GeV^{-2}$ for events with proton dissociation. 
Altogether one finds:
\be
\frac{d\sigma}{dt}(\gamma)\;\approx\;7.5\; nb \;\exp(-5.3\; |t|\; GeV^{-2})
\ee
Taking the measured cross section for $J/\Psi p$ at the same energy 
and divide it by 9 we find a value of similar size. The main difference
is the much smaller slope of $1.6~GeV^{-2}$ which is
due to the large mass:
\be
\frac{d\sigma}{dt}(J/\Psi)\;\approx\;4.4 \;nb \;\exp(-1.6\; |t|\; GeV^{-2})
\ee
for $W=200GeV$.
\begin{figure}[t]
\begin{center}
\epsfig{file=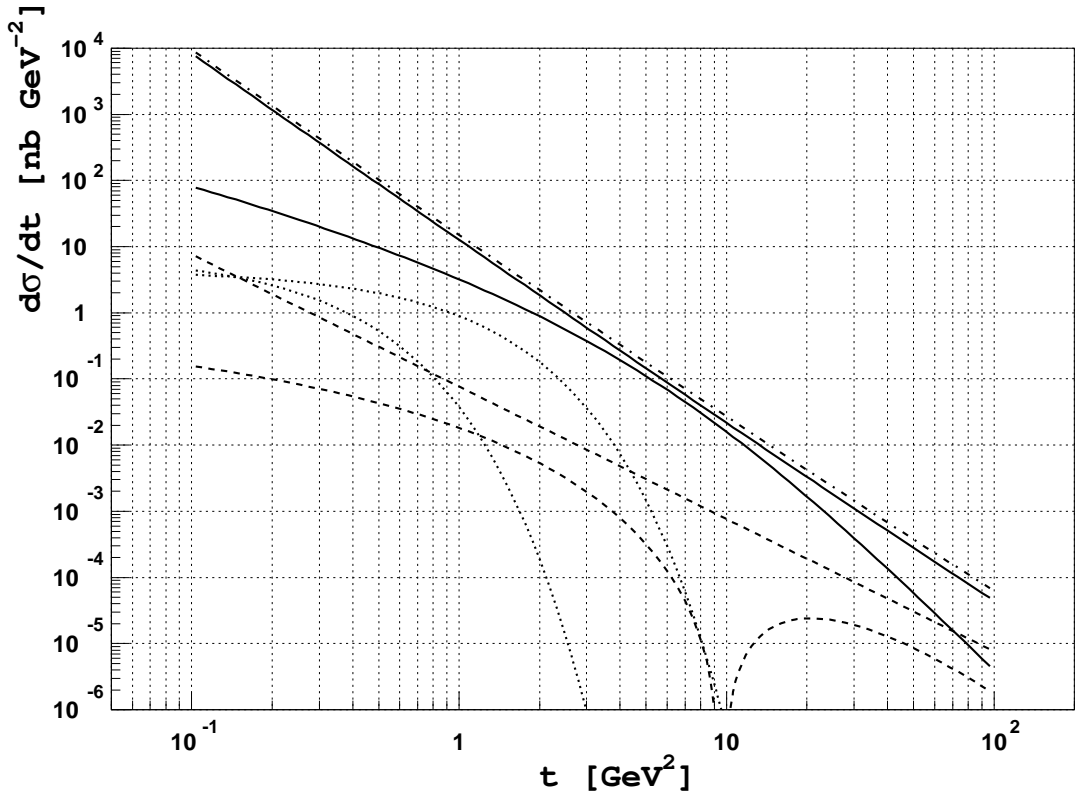,height=12cm}
\caption{The cross section for diffractive production of photons
and $J/\Psi$'s at $W=200~GeV$. 
The solid line denotes the BFKL-solution (upper line
for photons), the dashed line the leading order two-gluon exchange
(upper line again for photons) and the dotted line shows the VDM-estimate
(in this case the upper line is for $J/\Psi$). In addition the saddle point 
approximation is given as dash-dotted line.}
\end{center}
\end{figure}

In Fig.2 we show all discussed options in one plot. For the perturbative
result we have assumed a constant value for the strong coupling of 
$\alpha_s=0.2$ and $W=200~GeV$. 
The two solid lines are related to the full BFKL-solution,
the upper line denotes the production of a photon and the lower line 
the production of $J/\Psi$. In addition to the upper solid line we have plotted
the saddle point solution (\ref{saddlenonflip}) and (\ref{saddleflip})
for $\gamma q$-scattering which lies almost on top of the solid line.
The dashed lines show in a similar fashion the cross sections based
on the leading order two-gluon exchange. Again the upper line denotes
the production of a photon and the lower the production of $J/\Psi$.
The strongly curved, dotted lines represent the VDM-estimates. In this case
the line related to the production of $J/\Psi$ lies above the line for
the production of a photon due to the smaller $t$-slope.

The VDM-result for $\gamma q$-scattering
seems to contradict the perturbative result.
Moreover, a matching between the low $t$ nonperturbative and high $t$
perturbative regime seems to be difficult. For $J/\Psi$ on the other hand
a matching seems feasible. The enhancement due to BFKL is extremely large,
in both cases it is a factor of about 100 at 
$|t|=3~GeV^2$ and somewhat smaller (a factor 10) at $|t|=100~GeV^2$.

The fact that the perturbative BFKL-result for $\gamma q$-scattering
overshoots the VDM-result
so massively makes it hard to believe that the perturbative prediction
is close to the true value. There are two main reasons
to mention. First, the BFKL-solution is implemented at Leading Order.
NLO-corrections are known to reduce the cross section substantially.
Second, although we consider large-$t$ the internal integration over the
transverse momenta in the virtual loops is performed without
any infrared cutoff. The result is still finite, but dominant
contributions might come from the infrared region and thus is influenced
by confinement. This point needs further investigation by imposing
an infrared cutoff on the separation of the quark-antiquark pair.
The consequence of all conceivable corrections might be a reduction
of the cross section by 1 or 2 orders of magnitude.

\section{Conclusions}
We have calculated the cross section for $\gamma q$-elastic 
scattering in the Regge limit ($W^2\gg |t|\gg \Lambda^2_{QCD}$). 
The BFKL-solution for nonzero momentum transfer was
used leading to a strong enhancement of the cross section.
Believing in this prediction a measurement of the process should
be feasible. The comparison with a VDM-calculation indicates, however,
that the perturbative prediction might be much to high. The simple fact of
observing events at HERA or not will already give a hint with regard to 
the validity of the present pQCD result.
We also found that the cross section for elastic $\gamma q$-scattering
exceeds the cross section for $\gamma q\rightarrow J/\Psi q$.

Some effort was put in the detailed presentation of the method
employed for performing the convolution of the photon wave function with the
conformal eigenfunction. The integrals we were faced with are typical for two
dimensional conformal field theories. They seem to be solved most 
efficiently when complex variables are introduced and the integration is
factorized in complex times complex-conjugate contributions. A similar
technique was used in ref.\cite{Kor}.
We have pointed out that special attention has to be given 
to the problem of large infrared contributions. One way of tackling this
problem is to keep the virtuality $Q^2$ of the initial photon high enough 
\cite{ForEv}. It will be interesting to see how much the 
$\gamma^* q$-cross section will decrease 
when the photon virtuality is increased from 0 to $1~GeV^2$.

\section*{Acknowledgements}
We are very grateful for the intense discussion with J. Forshaw 
and N. Evanson. N. Evanson was also so kind
to point out missprints in our draft.
Special thanks go to M. Ryskin for valuable discussions on the problem
of infrared contributions. 
D.Yu I. would like to acknowledge the warm hospitality 
extended to him at University of Leipzig.  
This project was supported in parts by the RFBR under 
contract No 960219079, the Deutsche
Forschungsgemeinschaft under contract No Schi 422/1-2 and by the EU
Fourth Framework Programme  
``Training and Mobility of Researchers''  Network,    
``Quantum Chromodynamics and the Deep Structure of Elementary
Particles'', contract   FMRX-CT98-0194 (DG~12-MIHT).

\end{document}